\documentclass[preprint,12pt,authoryear]{elsarticle}

\usepackage{amssymb}

\usepackage{lineno}

\usepackage[version=3]{mhchem}
\makeatletter
\newcounter{reaction}
\renewcommand\thereaction{C\,\arabic{reaction}}
\newcommand\reactiontag{\refstepcounter{reaction}\tag{\thereaction}}
\newcommand\reaction@[2][]{\begin{equation}\ce{#2}%
\ifx\@empty#1\@empty\else\label{#1}\fi%
\reactiontag\end{equation}}
\newcommand\reaction@nonumber[1]{\begin{equation*}\ce{#1}%
\end{equation*}}
\newcommand\reaction{\@ifstar{\reaction@nonumber}{\reaction@}}
\makeatother

\journal{Astrobiology}

\begin{document}

\begin{frontmatter}



\title{Hypotheses for near-surface exchange of methane on Mars\tnoteref{label0}}

\tnotetext[label0]{Copyright 2016 California Institute of Technology. U.S. Government sponsorship acknowledged.}

\author[label1,label2]{Renyu Hu\corref{cor1}}
\author[label1]{A. Anthony Bloom}
\author[label2]{Peter Gao}
\author[label1]{Charles E. Miller}
\author[label1,label2]{Yuk L. Yung}

\address[label1]{Jet Propulsion Laboratory, California Institute of Technology, Pasadena, CA 91109, USA}
\address[label2]{Division of Geological and Planetary Sciences, California Institute of Technology, Pasadena, CA 91125, USA}

\cortext[cor1]{renyu.hu@jpl.nasa.gov}

\begin{abstract}
The Curiosity rover recently detected a background of 0.7 ppb and spikes of 7 ppb of methane on Mars. This in situ measurement reorients our understanding of the Martian environment and its potential for life, as the current theories do not entail any geological source or sink of methane that varies sub-annually. In particular, the 10-fold elevation during the southern winter indicates episodic sources of methane that are yet to be discovered. Here we suggest a near-surface reservoir could explain this variability. Using the temperature and humidity measurements from the rover, we find that perchlorate salts in the regolith deliquesce to form liquid solutions, and deliquescence progresses to deeper subsurface in the season of the methane spikes. We therefore formulate the following three testable hypotheses. The first scenario is that the regolith in Gale Crater adsorbs methane when dry and releases this methane to the atmosphere upon deliquescence. The adsorption energy needs to be 36 kJ mol$^{-1}$ to explain the magnitude of the methane spikes, higher than existing laboratory measurements. The second scenario is that microorganisms convert organic matter in the soil to methane when they are in liquid solutions. This scenario does not require regolith adsorption, but entails extant life on Mars. The third scenario is that deep subsurface aquifers produce the bursts of methane. Continued {\it in situ} measurements of methane and water, as well as laboratory studies of adsorption and deliquescence, will test these hypotheses and inform the existence of the near-surface reservoir and its exchange with the atmosphere.
\end{abstract}


\end{frontmatter}


\section{Introduction}

Methane (\ce{CH4}) is an organic molecule in Earth's atmosphere primarily produced by living organisms \citep{Seinfeld2006}. It has also been measured in Mars's atmosphere by telescopic and spacecraft remote sensing \citep{Formisano2004,Krasnopolsky2004,Mumma2009}. However, these measurements produced inconsistent results and some of these measurements have been called into question \citep[e.g.][]{Zahnle2011}. Recently, an {\it in situ} measurement of methane on Mars has been made: MSL's Tunable Laser Spectrometer (TLS) as a part of the Sample Analysis at Mars (SAM) instrument determined a background \ce{CH4} mixing ratio of $\sim0.7$ ppbv and a pulse of $\sim7$ ppbv observed over two months \citep{Webster2015}. These measurements suggest strong temporal variability of the methane abundance on the surface of Mars.

Methane's atmospheric existence requires a geologically recent or continually replenishing source, because methane has an lifetime of $\sim300$ years in Mars' oxidizing atmosphere \citep{Nair1994,Summers2002}. This source can be photo-degradation of organic matter in the meteorites fallen onto Mars \citep{Keppler2012}.  Alternatively, methane has to come from Mars itself, which challenges the conventional understanding of a geologically and biologically dead Mars \citep[e.g.,][]{Lyons2005,Atreya2007}. An indigenous source of methane is corroborated by a recent discovery of methane evolved from Martian meteorites \citep{Blamey2015}. Furthermore, atmospheric processes alone cannot produce the variability of methane detected by the Curiosity, because the atmospheric mixing time of methane is much shorter than its chemical lifetime \citep{Lefevre2009}. The discovery of methane thus compels a new chapter of Mars research to explain the existence of methane and its variability in the Martian atmosphere.

Extrapolating our knowledge of terrestrial biotic sources to Mars, many consider methanogens (a type of \ce{CH4}-producing Archaean microbe) as a probable analog to Martian life forms \citep{Boston1992,Weiss2000,Schulze2008}. Some methanogens are able to utilize inorganic compounds (\ce{H2} and \ce{CO2}) as their only source of energy and produce methane. Being independent of photosynthesis for subsistence, methanogens can thrive in the deep subsurface where \ce{CO2} is the predominant oxidant and \ce{H2} (aq) is abundant from water-rock interactions (e.g., ferrous-iron reduction of \ce{H2O} to \ce{H2} during serpentinization \citep{Lyons2005,Chassefiere2011}). \ce{H2} may also come from photolysis of \ce{H2O} in the atmosphere \citep{Weiss2000}. In fact, methanogens thrive in some of the harshest environments on Earth, including extremely acidic environments and inside Greenland glacial ice 3-km deep, which is analogous to Martian subsurface ice environments \citep{Stevens1995,Chapelle2002,Walker2005,Tung2005}.

Alternatively, Fischer-Tropsch-type (FTT) reactions can be a potential methane source \citep{Oze2005,Atreya2007,Etiope2013}. FTT is the most widely posited abiotic source of methane on Earth. Catalyzed by transition metals and related oxides, these reactions have the same overall chemistry as the methanogenesis, take place in hydrothermal environments, and probably source \ce{H2} also from serpentinization. Abundant evidence indicates that volcanism and hydrothermal environments existed, and might still exist, on Mars \citep{Schulze2008,Squyres2008,Hauber2011}. These environments provide heat and liquid water to support FTT and/or microbial \ce{CH4} production. Thus, atmospheric \ce{CH4} may point to either serpentinization or the existence of life itself, both of which are associated with a warm, habitable backdrop. 

In this paper we focus on the singularly important discovery of the episodically enhanced \ce{CH4} emission on Mars. The elevated methane levels occurred in the southern-hemisphere winter, except for a single and statistically marginal measurement on Sol 306 \citep{Webster2015}. 

A first-order question about the Curiosity's methane spikes is whether they represent new source to the atmosphere, or some cyclic processes that conserve methane. If the methane spikes represent new methane produced from the deep subsurface \citep{Atreya2007} or from a meteoritic source \citep{Fries2015}, then this methane would be oxidized and remain in the Martian atmosphere in the form of \ce{CO2}. As a lower estimate, if 7 ppb of methane is produced per year across the entire Martian surface (representing the spike from ${\rm L}_{\rm s}\sim50^{\circ}-90^{\circ}$), then over 3 billion years, the total amount of \ce{CO2} produced is 20 times the present atmospheric \ce{CO2} content. This is much higher than the Amazonian outgassing rates determined from photogeologic constraints \citep{Greeley1991, Grott2011} or what would be allowed by atmospheric evolution models \citep{Chassefiere2011,Hu2015}. On the other hand, if the methane is converted to organics and stored in the regolith, then the molar density of organics originating from methane would be (8.4/$d$) mol per gram of soil, assuming \ce{C6H5Cl} as the organic molecule, with $d$ as the depth in the soil in units of cm to which the organics are stored. This is far greater than the observed molar density for any reasonable values of $d$ \citep{Freissinet2015}. Therefore, if the spikes are new methane, the source must be local; alternatively, the spikes are signatures of cyclic processes that produce fast source and sink.


Interestingly, the methane spikes were coincident with surface relative humidities greater than 60\% measured by the Rover Environmental Monitoring Station (REMS) \citep{Webster2015}. This apparent correlation motivates us to consider the surface-atmosphere exchange as a major modulator for the atmospheric methane abundances in Gale Crater, the landing site of MSL. 

A high surface relative humidity may have a strong impact on the atmosphere-surface exchange because perchlorate salts in the regolith may deliquesce to form liquid solutions \citep{MartinTorres2015}. Martian regolith contains $\sim$0.5\% perchlorate salts by weight, measured at both the Phoenix and the Curiosity landing sites \citep{Hecht2009,Leshin2013,Ming2014}. They can form liquid solutions under Martian conditions due to their low eutectic temperatures and low deliquescence relative humidities (DRH) \citep{Marion2010, Martinez2013, Toner2014, Nuding2014, Fischer2014, MartinTorres2015}. Recent laboratory measurements have determined that the DRH of calcium perchlorate is only $\sim50\%$ at 198 K, and that the wet salt does not lose moisture until the relative humidity drops to $\sim15\%$ \citep{Nuding2014}. By comparing the eutectic temperature and the DRH to the measured surface temperature and relative humidity, we confirm that perchlorate salts in the subsurface of Gale Crater may deliquesce in the southern winter, in which the methane spikes were measured.

We therefore formulate three hypotheses in an attempt to explain the variability of the atmospheric methane abundance at Gale Crater. The first hypothesis (Hypothesis I) is that the regolith in Gale Crater adsorbs methane when dry and releases this methane to the atmosphere when the relative humidity in the regolith is high enough for perchlorate salts to deliquesce. The second hypothesis (Hypothesis II) is that microorganisms convert organic matter in the soil to methane when they are in liquid solutions. This scenario does not require regolith adsorption. The third hypothesis (Hypothesis III) is that deep subsurface aquifers produce the bursts of methane. The first two hypotheses explain the methane spikes as signatures of cyclic processes that produce fast source and sink, whereas the third hypothesis explains the methane spikes as local emission of new methane.

\section{Subsurface temperature and humidity models}

We model the subsurface temperature and humidity of dusty terrains at Gale Crater by solving diffusion equations of heat and moisture with the hourly-averaged REMS data as the boundary conditions \citep{Gomez-Elvira2012}. The absolute humidity of each measurement is computed from the measured relative humidity and the sensor temperature archived in NASA's Planetary Data System (PDS; https://pds.nasa.gov/), with the saturation vapor pressure from \cite{Murphy2005}. The uncertainty of each surface temperature measurement is $\sim5$K at low temperatures \citep{Gomez-Elvira2014}, and the uncertainty of each relative humidity measurement is $<2$\% \citep{Harri2014}. The combined uncertainty of the hourly-averaged values is one-order-of-magnitude smaller for Gaussian statistics. The data clearly shows the effect of the rover traversing diverse terrains, as the magnitudes of the diurnal variations in both temperature and humidity change significantly \citep{Gomez-Elvira2014, Harri2014}. To remove this effect, we use daily means for a Mars year (from Sol 39 to Sol 707) and add the hourly means from Sol 70 to Sol 89 (when the rover is at Rocknest, a dusty terrain) to the daily means. We use the diurnal variation measured over a dusty terrain as a proxy for regolith diurnal temperature and moisture variability. We neglect the seasonal change in the magnitude of the diurnal variation in temperature and humidity, which is valid given the equatorial location of Gale Crater.

\begin{figure}[h]
\begin{center}
 \includegraphics[width=0.8\textwidth]{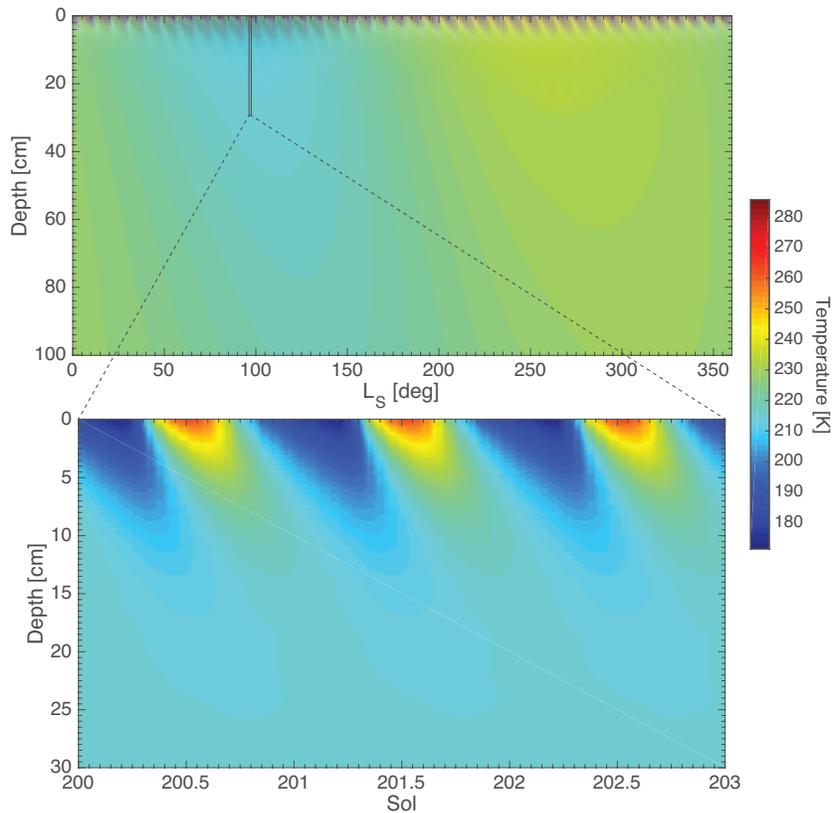}
 \caption{
Subsurface temperature derived from REMS surface temperature measurements. The bottom panel provides a zoom-in view of the top panel over 3 Sols. Diurnal temperature variation is limited to a depth of a few cm, and seasonal temperature variation extends to $\sim1$ m.}
 \label{Temp}
  \end{center}
\end{figure}

The REMS relative humidity data in the daytime are thought to be unreliable, because the daytime relative humidity is often lower than 2\% \citep{Harri2014}. Multiplied by the saturation vapor concentration, the REMS daytime relative humidity measurements imply atmospheric water concentrations up to $3\times10^{-5}$ kg m$^{-3}$. This water concentration appears to be too large compared with the column water abundances in the equatorial region. The Mini-TES instrument on Spirit and Opportunity rovers measured daytime water columns seasonally variable from a few to 20 pr. $\mu$m \citep{Smith2006}. The TES instrument onboard Mars Global Surveyor also indicated a peak daytime water column of 20 pr. $\mu$m \citep{Smith2004}. Using a boundary layer thickness of 5 km \citep{Savijarvi2015}, this column corresponds to a concentration of $4\times10^{-6}$ kg m$^{-3}$ at the surface. If the boundary layer has a thickness of 2 km, the surface concentration can be up to $1\times10^{-5}$ kg m$^{-3}$. However, what if the REMS daytime data are actually right? A detailed inspection reveals that the several hundred measurements in each hour in the daytime are not consistent with a random noise around zero. The standard deviation of each group of measurements is small enough to statistically declare a non-zero mean value at $10-\sigma$ or higher. Of course, the standard deviation may be artificially small as a result of the data processing pipeline. We therefore consider all these possibilities of daytime water concentration in the analysis: for case L (low water concentration) we impose an upper limit of $4\times10^{-6}$ kg m$^{-3}$ on the REMS data; for case M (intermediate water concentration) we impose an upper limit of $1\times10^{-5}$ kg m$^{-3}$; and for case H (high water concentration) we use the REMS data as reported in the PDS.

\begin{figure}[h]
\begin{center}
 \includegraphics[width=1.0\textwidth]{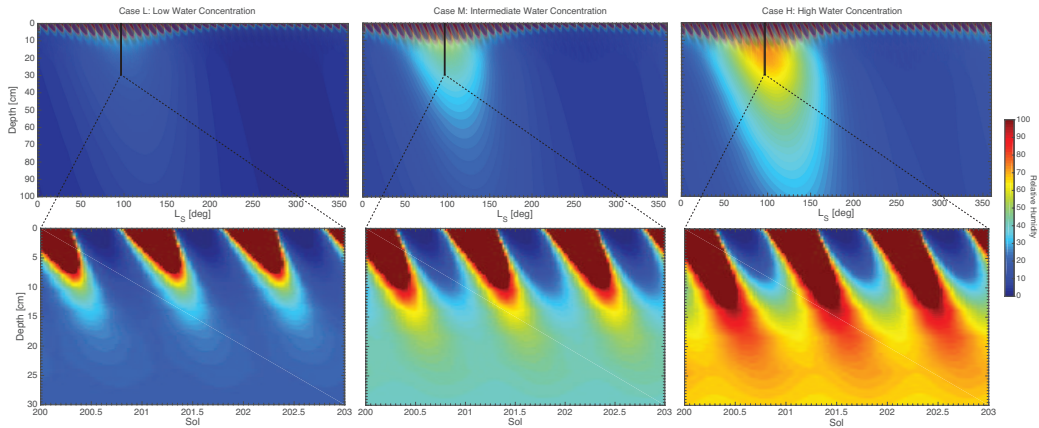}
 \caption{
Subsurface relative humidity derived from REMS surface temperature and near-surface relative humidity measurements. The bottom panels provide zoom-in views of the top panels over 3 Sols. The three columns correspond to three cases for daytime water concentration. }
 \label{Moist}
  \end{center}
\end{figure}

The thermal and moisture diffusion equations are solved on a 41-level grid from zero to 5.2 m (i.e., four thermal skin depths). The size of layers increases with depth, and the smallest size is 0.01 cm at the shallow subsurface. We use the thermal inertia derived for Rocknest \citep{Martinez2014}, and the volumetric heat capacity for the Martian aeolian dunes \citep{Edgett1991}. Figure \ref{Temp} shows the modeled subsurface temperature. The moisture diffusion equation (i.e., Eq. 13 in \cite{Zent1993}) includes physical adsorption and desorption by regolith, condensation, and corrections due to soil porosity and tortuosity \citep{Mellon1993,Zent1993}. We assume a soil porosity of 0.5, and a porosity/tortuosity ratio of 0.3  \citep{Sizemore2008}. Physical adsorption and desorption is assumed to be in equilibrium, and follows a Langmuir-Freundlich isotherm measured for palagonite, a terrestrial analog of Martian basaltic soil \citep{Zent1997}. The effect of physical adsorption and desorption is to retard water vapor diffusion by several orders of magnitude \citep{Fanale1971,Zent1997}. Figure \ref{Moist} shows the model subsurface relative humidity for three assumptions of the daytime water concentration. 

The diurnal variation of relative humidity extends to a depth of $\sim10$ cm, on the same order of magnitude as the diurnal temperature variation. Below the depth of diurnal variation, the relative humidity is constantly below 20\%, except for the southern winter during which the relative humidty can be as high as 50\% within the depth of 1 m for the high water concentration scenario (Figure \ref{Moist}). Further inspection of the results reveals that the absolute humidty below the depth of diurnal variation actually has little seasonal variation, consistent with the picture that water vapor diffusion is significantly retarded by physical adsorption. Therefore, the seasonally high subsurface relative humdity is driven by the seasonally low subsurface temperature as shown in Figure \ref{Temp}. Furthermore, a zoom-in view of the relative humidity variation indicates that condensation can occur near the surface early morning during the southern winter (Figure \ref{Moist}), a finding consistent with \cite{MartinTorres2015}.

\begin{figure}[h]
\begin{center}
 \includegraphics[width=0.5\textwidth]{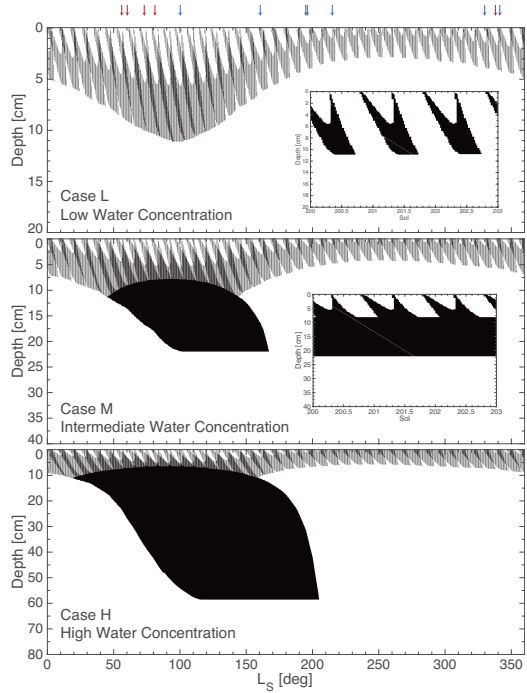}
 \caption{
Subsurface layers in which deliquescent perchlorate solution is expected. The depth range in which calcium perchlorate salts are deliquescent and forming liquid brine is shown in black color. The three panels correspond to three cases for daytime water concentration. Note different vertical scales among the panels. 
The two inserts provide zoom-in views of three Sols to illustrate the diurnal cycle of deliquescence and efflorescence in the shallow subsurface.
The arrows on the top indicate the ${\rm L_S}$ when SAM's methane measurements were taken, with the blue arrows indicating background measurements and the red arrows indicating the spike measurements.
}
 \label{Deli}
  \end{center}
\end{figure}

The modeled temperature and relative humidity are then compared with the experimentally measured DRH for perchlorate \citep{Nuding2014} to determine whether deliquescence occurs (Figure \ref{Deli}). We assume calcium perchlorate because both Phoenix and Curiosity identify calcium perchlorate as the likely form of the parent salt \citep{Glavin2013,Kounaves2014}. We find that the surface temperature and humidity conditions allow deliquescence to occur in the top $5-15$ cm of soil, each Sol before sunrise and after sunset (Figure \ref{Deli}). The liquid solution does not persist over diurnal cycles close to the surface. The depth range of the transient liquid solution varies with the surface daily mean temperature and is largest in the southern winter.

For a conservative daytime water concentration of $4\times10^{-6}$ kg m$^{-3}$ or lower (Case L), deliquescence only occurs in the shallow subsurface and results in transient solutions. This is consistent with the findings of \cite{MartinTorres2015}, in which they omitted all daytime relative humidity measurements. Nonetheless, if the daytime water concentration is indeed higher, our calculation indicates that persisting liquid solutions could occur below the shallow subsurface in the southern winter, starting from ${\rm L}_{\rm S}\sim50^{\circ}$ \footnote{The solar longitude ${\rm L}_{\rm S}$ is the Mars-Sun angle, measured from the Northern Hemisphere spring equinox where ${\rm L}_{\rm S}=0$. For the southern hemisphere, ${\rm L}_{\rm S}=90$ corresponds to winter solstice, and ${\rm L}_{\rm S}=270$ corresponds to summer solstice.} (Figure \ref{Deli}, Cases M and H). The range of the liquid solution is larger for a higher water concentration. Because of a small diffusion coefficient, the absolute humidity of the soil at depths is roughly constant. The deep liquid solution is formed by slow cooling of the deep soil and the increase of the relative humidity. The liquid solution persists to ${\rm L}_{\rm S}\sim180-210^{\circ}$, longer than the high relative humidity season, manifesting hysteresis.

\section{Hypothesis I: deliquescence-modulated adsorption and desorption of methane}

This scenario assumes that methane is normally adsorbed by the regolith. However, when deliquescence occurs, the liquid solutions of perchlorate that form can conceivably coat the soil particles and deactivate most active sites, releasing methane into the atmosphere. Wetted particles have much smaller surface area than dry particles. 10-$\mu$m round-shape particles of basaltic composition only have surface area of $\sim0.15$ m$^2$ g$^{-1}$, and this is even smaller for larger particles. Dry regolith on Mars, however, has surface area $\sim17-100$ m$^2$ g$^{-1}$ \citep{Ballou1978,Zent1997,Meslin2011}. It is therefore plausible to hypothesize that wetting could effectively remove surface area and displace adsorbed methane. The methane is then mixed in the planetary boundary layer, and may be transported and re-adsorbed by the regolith elsewhere. 

Taking the boundary layer as a box (e.g., 33\% of the column mass for a boundary layer thickness of 5 km), the column mass of methane in the box ($M_{\ce{CH4}}$) is modeled by the following equation,
\begin{equation}
\frac{dM_{\ce{CH4}}}{dt} = F_{\rm Background} - \frac{M_{\ce{CH4}}}{t_{\rm res}} + F_{\rm Deliquescence}. \label{ad}
\end{equation} 
$t_{\rm res}$ is the local atmospheric residence time of methane that brackets removal of methane by diurnal thermal tides, horizontal advection and diffusion, and soil adsorption. The horizontal advection and diffusion can be a sink in this model because Gale Crater is not fully covered with thick regolith, so spreading methane emitted from regions of thick regolith across Gale Crater can greatly reduce the local methane concentration. $t_{\rm res}$ is taken as a free parameter, because the footprint of the methane emission is unknown. $F_{\rm Background}$ is the flux of methane to maintain the 0.7-ppb background level of methane, and is equal to $(M_{\ce{CH4}})_{\rm Background}/t_{\rm res}$. $F_{\rm Background}$ is invoked to complete the equation and it represents the partitioning of methane between the atmosphere and the regolith. Equation (\ref{ad}) can be rewritten using the departure from the background level $M_{\ce{CH4}}-(M_{\ce{CH4}})_{\rm Background}$ as the variable and without the $F_{\rm Background}$ term.

Once deliquescence occurs, the displaced methane is assumed to be released to the atmosphere immediately. This is probably close enough to reality because the deliquescence starts from the shallow subsurface and progresses deeper, because solubility of methane in salty water is negligible, and because the diffusivity of methane is fast enough such that it should quickly reach the surface \citep{Gough2010}. The flux of methane produced by the onset of deliquescence is
\begin{equation}
F_{\rm Deliquescence} = \frac{d}{dt}\int_{\rm Deliquescence} \rho_s \gamma_{\ce{CH4}} m_{\ce{CH4}} A_s\theta_{\ce{CH4}} dz, \label{dflux}
\end{equation}
where $\rho_s\sim1300$ kg m$^{-3}$ is soil density, $\gamma_{\ce{CH4}}=5.2\times10^{18}$ m$^{-2}$ is the monolayer coverage of methane per unit surface area \citep{Gough2010}, $m_{\ce{CH4}}$ is the mass of methane molecule, $\theta_{\ce{CH4}}$ is the coverage ratio of the methane, and $A_s\sim17-100$ m$^2$ g$^{-1}$ is the specific surface area. The specific surface area adopted in the study is estimated based on in situ measurements of Viking \citep{Ballou1978}, or measurements of JSC Mars-1 analog \citep{Meslin2011}. The specific surface area could be as high as 1000 m$^2$ g$^{-1}$ if the soil was mainly made of clay minerals \citep{Zent1997}, which is not seen in Gale Crater.

$\theta_{\ce{CH4}}$ is calculated from the Langmuir isotherm,
\begin{equation}
\theta_{\ce{CH4}} = \frac{K_{eq}n_{\ce{CH4}}}{1+K_{eq}n_{\ce{CH4}}+K'_{eq}n_{\ce{CO2}}}, \label{langmuir}
\end{equation}
where $n_{\ce{CH4}}$ is the number density of methane in the gas phase, and
\begin{equation}
K_{eq} = \frac{vh}{4\gamma_{\ce{CH4}}k_bT}\exp{(E_a/RT)},
\end{equation}
where $v$ is the thermal velocity of methane, $h$ is the Planck constant, $k_b$ is the Boltzmann constant, $R$ is the gas constant, and $E_a$ is the adsorption energy. $n_{\ce{CO2}}$ is the number density of \ce{CO2} and $K'_{eq}$ is the equilibrium constant for \ce{CO2} adsorption. The Equation (\ref{langmuir}) is applicable with \ce{CO2} being the sole competing substance for adsorption, because \ce{CO2} ice cannot form in Gale Crater, \ce{H2O} ice may form but only at the top $5\sim10$ cm of soil during nighttime, and inhomogeneous grains, like perchlorate grains, only make less than 1\% of soil mass. Given that in any case $K_{eq}n_{\ce{CH4}}\ll1$, Equation (\ref{langmuir}) is reduced to
\begin{equation}
\theta_{\ce{CH4}} = K_{eq}n_{\ce{CH4}}(1-\theta_{\ce{CO2}}),
\label{langmuirr}
\end{equation}
where $\theta_{\ce{CO2}}$ is the coverage ratio of \ce{CO2},
\begin{equation}
\theta_{\ce{CO2}} = \frac{K'_{eq}n_{\ce{CO2}}}{1+K'_{eq}n_{\ce{CO2}}}. \label{langmuirc}
\end{equation}
We adopt the $\theta_{\ce{CO2}}$ measurement for palagonite under Martian conditions fitted to the format of Equation (\ref{langmuirc}) by \cite{Zent1995}.

$E_a$ has been measured over the Martian soil analog JSC-Mars-1 and the measured value is $18\pm2$ kJ mol$^{-1}$ \citep{Gough2010}. We however postulate that the detail surface properties could matter for the amount of methane adsorption, specified by $E_a$ in our formulation. To account for model uncertainties, including surface properties, advection, diffusion and the effect of topography on diurnal \ce{CH4} accumulation, we treat $E_a$ and $t_{\rm res}$ as free parameters.

\begin{figure}[h]
\begin{center}
 \includegraphics[width=0.8\textwidth]{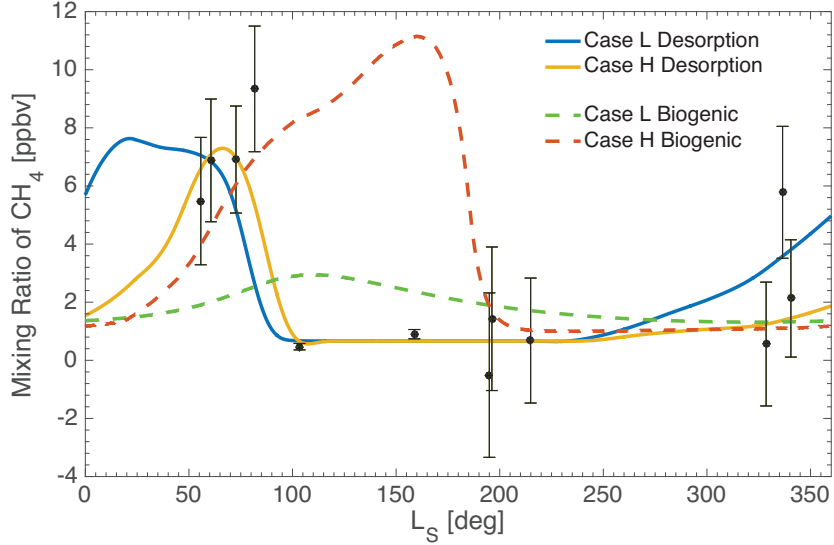}
 \caption{
Models of methane variability due to deliquescence-induced adsorption and desorption (Hypothesis I) or deliquescence-enabled methanogenesis (Hypothesis II) in comparison with MSL measurements \citep{Webster2015}. 
For Hypothesis I, the best-fit model of Case L (low water concentration) requires an adsorption energy of 37.2 kJ mol$^{-1}$, and a residence time of 12 days. The goodness of fit is $\chi^2/dof=2.2$. The best-fit model of Case H (high water concentration) requires an adsorption energy of 36.4 kJ mol$^{-1}$, and a residence time of 4 days. The goodness of fit is $\chi^2/dof=1.4$. 
For Hypothesis II, none of the models produce a good fit because of the high flux at ${\rm L}_{\rm S}\sim100^{\circ}$. We instead show two examples for illustration: both cases assume a baseline residence time of soil carbon of 30 days; the case L assumes $t_{\rm res}$ of 100 Sols and the case H assumes $t_{\rm res}$ of 10 Sols.
 }
 \label{fit1}
  \end{center}
\end{figure}

We solve Equations (\ref{ad}-\ref{dflux}) to determine whether methane adsorption and desorption would contribute to the variability seen in Gale Crater. Figure \ref{fit1} shows the best-fit models for Case L (low water concentration) and Case H (high water concentration). We only include the day-to-day change in the range of deliquescence when evaluating Equation (\ref{dflux}). This is because adsorption of methane is kinetically slow \citep{Gough2010} and a diurnal deliquescence-efflorescence cycle would not allow the soil to be loaded by methane when dry. We hypothesize that the seasonal variation in the maximum depth of deliquescent soil (Figure \ref{Deli}) could be the main driver of methane variability in the atmosphere. 

Figure \ref{fit1} shows that deliquescence-induced desorption of methane can account for the methane spikes. The season in which deliquescence progresses to deeper subsurface, in both low and high water cases, is consistent with the season in which the methane spikes are measured (i.e., the beginning of southern winter). The methane released from the regolith can quickly dissipate, consistent with the rapid drop-off at ${\rm L}_{\rm S}\sim100^{\circ}$, for a residence time of a few days. A small residence time indicates a small footprint of the methane source in Gale Crater, and fast re-adsorption of methane to the regolith. This model does not predict a methane spike at ${\rm L}_{\rm S}\sim330^{\circ}$. However, given the low significance of the measurement, the overall goodness of fit achieved by this hypothesis ($\chi^2/dof\sim1.4$) is satisfactory. 

The main challenge of this model is that the adsorption energy required is 2-fold greater than what is measured by \cite{Gough2010}. This is because a sufficient amount of methane needs to be temporarily stored in the upper 10 cm (Case L) or 60 cm (Case H) of the soil layer to account for the 7-ppb spike. Increasing the residence time (e.g., enlarging the footprint) could help reduce the adsorption energy, but the residence time is limited by the firm non-detection at ${\rm L}_{\rm S}\sim100^{\circ}$. How is the required adsorption energy compared with that of methane for other materials? Existing measurements, often performed under room or higher temperatures, suggest a fairly wide range of adsorption energy: 15 - 25 kJ mol$^{-1}$ for various kinds of zeolites \citep{Zhang1991,Cavenati2004,Himeno2007}, 16 - 21 kJ mol$^{-1}$ for activated carbons \citep{Himeno2005}, and 39 kJ mol$^{-1}$ for a synthetic nano-porous titanium silicate \citep{Delgado2008}. Further laboratory studies are warrented to determine whether the high adsorption energy required by this scenario is possible for Martian regolith.



\section{Hypothesis II: biological conversion between organic matter and methane}

This scenario postulates extant microorganisms in the regolith, and assumes that in the presence of liquid water and organic compounds, these microorganisms are able to produce methane. Like Hypothesis I, this hypothesis is based on the seasonal deliquescence-efflorescence cycle, but it does not require a large adsorption energy. 

This hypothesis is motivated by the recent discovery of organic matter in drill samples of a mudstone in the Yellowknife Bay region of Gale Crater \citep{Freissinet2015}. The detected organic matter includes trace-level chlorobenzene (\ce{C6H5Cl}) and dichloroalkanes (\ce{C3H6Cl2}) in the evolved gases, whose precursors are suggested to be benzenecarboxylates and aliphatic hydrocarbons. Benzene rings are hard to break, but one may imagine microorganisms instead use the aliphatic hydrocarbons by performing the following hydrogenation reaction
\reaction{C3H8 + 2H2 -> 3CH4, \label{r2}}
which is exothermic ($\Delta_rG^{\circ}=-129.0$ kJ mol$^{-1}$). Alternatively, methanogens could directly source carbon from the atmosphere
\reaction{CO2 + 4H2 -> CH4 + 2H2O. \label{r1}}
The source of hydrogen for this metabolism is photochemical dissociation of water vapor in the atmosphere \citep{Weiss2000}. We imagine yeast-like microorganisms performing the biochemical reactions above in liquid solutions formed by deliquescence of perchlorate salts.



On Earth, methanogenesis occurs as a result of oxygen-depleted decomposition of organic C compounds, and consume acetate (\ce{CH3COOH}) to produce \ce{CH4} and \ce{CO2}, or \ce{CO2} and \ce{H2} to produce \ce{CH4} and \ce{H2O} \citep{Whalen2005}. The present-day production of \ce{CH4} predominantly occurs in water-logged, oxygen-depleted (anaerobic) environments. Temperature is a limiting factor to methanogenesis: \ce{CH4} emissions from soils have been found to vary exponentially as a function of temperature \citep{VanHulzen1999,Bloom2010,Yvon2014}. Other factors that influence the production of \ce{CH4} in oxygen-depleted environments include soil pH and redox potential \citep{Bloom2012}. 

Life is known to exist in highly salty and highly saline environments on Earth \citep{Boetius2009, Stueken2015}. However, it is doubtful that any terrestrial microorganisms, even those living in highly saline environments, can live in an eutectic brine of Ca perchlorate on Mars, because the water activity on Mars would be too low \citep{Rummel2014}. Phenomenologically, and as a bold assumption, we consider the possibility that Martian life has developed biological mechanism that allows functioning in such a harsh environment.

\begin{figure}[h]
\begin{center}
 \includegraphics[width=0.6\textwidth]{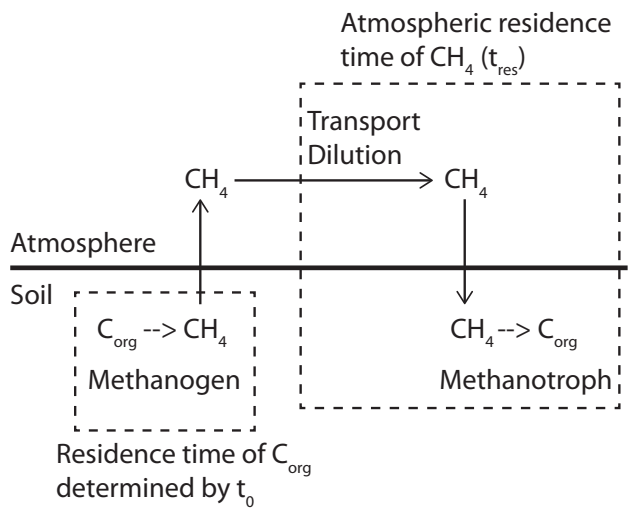}
 \caption{
Schematic illustration of time scales in Hypothesis II: biological conversion between organic matter and methane.
 }
 \label{bio_schematic}
  \end{center}
\end{figure}

Assuming methanogenesis occurs in Gale Crater, we model
\begin{equation}
\frac{dM_{\ce{CH4}}}{dt} = F_{\rm Background} - \frac{M_{\ce{CH4}}}{t_{\rm res}} + F_{\rm Methanogen}. \label{ad2}
\end{equation}
Here, $t_{\rm res}$ includes not only horizontal transport and physical adsorption, but also conversion from methane back to organic matter, probably by methanotrophs (see Figure \ref{bio_schematic}), given that \ce{CH4} oxidation in aerobic and anaerobic environments is a source of C and energy for terrestrial methanotrophs \citep{Whalen2005,Treude2007}. $F_{\rm Methanogen}$ is the methane flux produced by methanogens, which is modeled as
\begin{equation}
F_{\rm Methanogen} =  \frac{C \rho_s}{t_0}\int_{\rm Deliquescence} Q_{\rm 10}^{\frac{T-273.15}{10}} dz, \label{t_analog}
\end{equation}
where $C=5\times10^{-8}$ kg/kg is the soil content of aliphatic hydrocarbons, $t_0$ is the baseline residence time of organic carbon at 0 $^{\circ}$C, and $Q_{\rm 10}$ is an exponential temperature dependence constant. Equation (\ref{t_analog}) implies that $t_0$, together with the temperature and the liquid water availability in the soil, determines the residence time of organic carbon in the soil. The measured and modeled biological residence times of carbon in anaerobic environments span across several orders of magnitude (from 30 days for root exudates to $5\times10^5$ days for soil carbon \citep{Miyajima1997,Bridgham1998,Wania2010}). We choose $Q_{10}=2$, broadly consistent with terrestrial biological methane emission studies \citep{Bloom2010,Wania2013,Yvon2014}. 

The deliquescence-enabled methanogenesis can produce the methane spikes during the southern winter as measured by MSL, but cannot produce the rapid drop-off at ${\rm L}_{\rm S}\sim100^{\circ}$ (Figure \ref{fit1}). Compared with Hypothesis I, the methane spikes predicted by the methanogenesis model are later in the season. This is simply because the methane emission flux here is proportional to the thickness of the deliquescent soil, while the methane flux due to desorption is proportional to the derivative of that. For the same reason, this hypothesis is not directly consistent with the firm non-detection of elevated methane at ${\rm L}_{\rm S}\sim100^{\circ}$ or ${\rm L}_{\rm S}\sim150^{\circ}$, during which the subsurface soil remains wet. It is however possible that MSL was not downstream of the methane source during this period. 



\begin{figure}[h]
\begin{center}
 \includegraphics[width=0.6\textwidth]{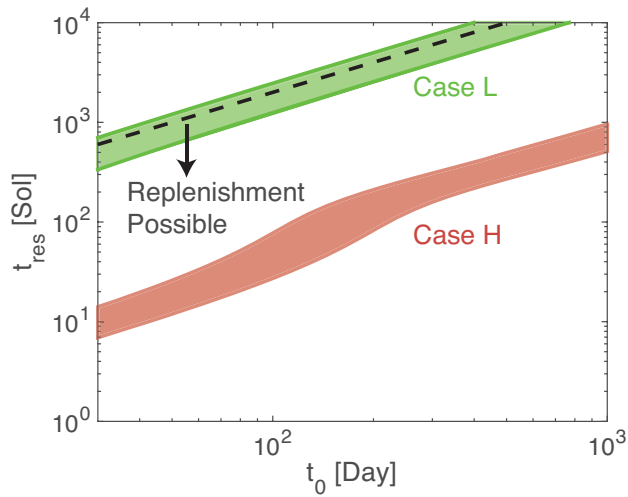}
 \caption{
Parameter ranges for the local atmospheric residence time of methane ($t_{\rm res}$) and the baseline residence time of soil carbon ($t_0$) that produce a $7\pm2$ ppb methane spike at ${\rm L}_{\rm S}\sim100^{\circ}$ as measured by MSL.
 }
 \label{fit2}
  \end{center}
\end{figure}

A high surface water concentration, and then a large thickness of deliquescent soil during winter is probably necessary for methanogenesis to produce the 7-ppb methane spikes. Figure \ref{fit2} shows the parameter ranges that produce the strength of the spikes. Additionally, one needs to consider replenishment of organic carbon in the soil. The methane production rate is controlled by the baseline residence time of soil carbon ($t_0$), which is a free parameter in the model. The atmospheric residence time of methane ($t_{\rm res}$) needs to be shorter than the lifetime of the organic carbon in the soil to ensure replenishment, if the main source of carbon of the methanogens is the soil organic carbon (Reaction \ref{r2}). In the model we find this condition corresponds to $t_{\rm res}<20-120\times t_0$, shown as a dashed line in Figure \ref{fit2}. Using this condition, we find that it is only marginally possible to produce the 7-ppb methane spikes in case L (low water concentration), while producing the methane spikes in case H (high water concentration) is much easier. If the methanogens directly source carbon from the CO$_2$ atmosphere (Reaction \ref{r1}), the atmospheric residence time of methane can be longer and the methane spikes are more readily fit by this scenario.

\section{Hypothesis III: Outbursts from subsurface permanent aquifer}

Another possibility is that the elevated methane measured by TLS represents sources of ``new'' methane emerging from a deep subsurface aquifer. The sources of methane may include subsurface gas-water-rock chemistry and microbial methanogenesis \citep{Lyons2005,Atreya2007}, with the difference being that biological methanogenesis is much faster than gas-water-rock reactions \citep{Yung2010}. Such a source of methane, along with cometary and meteoric sources, may be necessary to balance the loss of atmospheric methane due to UV photolysis over a time scale of a few hundred years. 

A subsurface aquifer may exist at a depth of 5 km assuming a geothermal gradient of 10 K/km \citep{Solomon1990,Clifford2010}. The aquifer can be partially sealed by an ice or clathrate layer to produce bursts of methane, in similar ways as the terrestrial arctic tundra. 
The arctic tundra is one of the major sources of methane for Earth's atmosphere. Concentrated bursts of methane have been observed at tundra sites in late fall as the seasonally thawed active layer refreezes, forcing sub-surface methane into the atmosphere \citep{Mastepanov2008}, and during the spring freeze-thaw transition when subsurface methane trapped by the frozen surface escapes \citep{Song2012}. The fall and spring bursts are transitory (occurring only for a $<10$-day window immediately surrounding the freeze-thaw transition) and episodic (they do not occur every season), and the magnitude of the emissions is highly variable. 

On Mars, however, if the methane was released deep from the subsurface, this release should be sporadic and have no seasonality. This is because the seasonal variations of temperature and relative humidity damp out at a depth of a few meters (Figure \ref{Temp}), and the condition of the permanent aquifer is not controlled by surface conditions. Other geological processes may trigger the subsurface methane to be vented out, such as seepage from mud volcanoes \citep[e.g.,][]{Etiope2011,Komatsu2011}.


The source of methane emission in this scenario must be localized, because on the mass balance calculation in \S~1. A source of methane close to MSL that emits $\sim7$ ppb of methane over a small area could be the cause of the methane spike, while the subsequent decrease back to the background level would be easily explained by the dispersal of the methane into the rest of the atmosphere. The small amount of methane released would have very little effect on the carbon budget of the rest of the atmosphere. It is an open question what areas of Mars would be amenable to such localized methane releases. 


\section{Conclusion and Prospects}

We propose the first working theories that explain the sub-annual variability of atmospheric methane abundance in Gale Crater, in the form of three testable hypotheses. First, adsorption of methane in the regolith up to a few tens cm depth, and its release driven by deliquescence during the winter, can explain the apparent methane variability. This scenario requires an adsorption energy 2-fold higher than laboratory measurements on Mars regolith analogs. Second, biological conversion from organic matter to methane in the regolith by microorganisms in perchlorate solutions can produce the magnitude of the methane spikes. However, this model cannot produce the rapid drop-off of the methane abundance at ${\rm L}_{\rm S}\sim100^{\circ}$, and likely requires a daytime water concentration higher than standard Mars atmospheric models. Third, the methane spikes may come from a deep subsurface aquifer, representing new methane to the atmosphere. Emitted intermittently, this methane may be the methane that maintains the 0.7-ppb background on a time scale of several hundred years, if the footprint of the emission is small.

Any of the three hypotheses, if confirmed, leads to profound ramification in our understanding of Mars as an active and potentially habitable world. If adsorption and desorption by regolith modulates atmospheric methane in Gale Crater, greater variability can be expected for higher latitudes, where perchlorate also exists \citep{Hecht2009} and deliquescence may occur for a larger fraction of the surface \citep{Martinez2013}. The methane variability suggested on the basis of telescopic observations \citep{Mumma2009}, though debated \citep{Zahnle2011}, could be explained in this way. Both the first and the second hypothesis work better when the daytime water concentration at the surface is high. If true, these scenarios warrant further investigation of the boundary layer dynamics and water vapor distribution on Mars. Lastly, if the methane spikes are from the deep subsurface, one might ask how lucky MSL has to be to observe the spikes and whether the geographic low of Gale Crater helps.

Continued monitoring of methane by MSL will test these hypotheses. The first two hypotheses predict the methane variability is seasonal and should repeat annually. The peak of the methane abundance occurs in different seasons between the two hypotheses, with the adsorption-desorption model in the early winter and the methanogenesis model in the later winter. We caution that observing the repeated signals, even if either of the hypotheses is true, can be hard because of the changing wind patterns and the rover location. The last hypothesis predicts the methane spikes are sporadic. Meanwhile, our investigations of the three hypotheses call for laboratory studies of gas adsorption and desorption during deliquescence, as well as further laboratory studies of the adsorption energy of methane for fine, porous silicate materials. Finally, future {\it in situ} Mars exploration may conduct improved characterization of the daytime water vapor concentration at the surface and measurement of the water's isotopic signatures will further constrain the rate of the atmosphere-regolith exchange, and reveal the smoking gun of a near-surface methane reservoir.

\section*{Acknowledgement}
Support for this work was partially provided by NASA through Hubble Fellowship grant \#51332.01 awarded by the Space Telescope Science Institute, which is operated by the Association of Universities for Research in Astronomy, Inc., for NASA, under contract NAS 5-26555.
YLY was supported in part by an NAI Virtual Planetary Laboratory grant
from the University of Washington to the Jet Propulsion Laboratory and
California Institute of Technology.
The research was carried out at the Jet Propulsion Laboratory, California Institute of Technology, under a contract with the National Aeronautics and Space Administration.



\section*{Reference}


\newcommand{\noopsort}[1]{} \newcommand{\printfirst}[2]{#1}
  \newcommand{\singleletter}[1]{#1} \newcommand{\switchargs}[2]{#2#1}

\end{document}